\documentclass[aps,prl,twocolumn,10pt]{revtex4-1}
\usepackage{graphicx,amsmath,amssymb}
\usepackage[utf8x]{inputenc}
\usepackage{url}
\usepackage{float}

\begin{document}

\title{Superresolution Microscopy of the Volume Phase Transition of pNIPAM Microgels}

\author{Gaurasundar M. Conley$^{1}$}
\author{Sofi N\"ojd$^{2}$}
\author{Marco Braibanti$^{1}$}
\author{Peter Schurtenberger$^{2}$}
\author{Frank Scheffold$^{1}$}
  \email{Frank.Scheffold@unifr.ch}

\affiliation{$^1$ Department of Physics, University of Fribourg, Chemin du Mus\'ee 3, 1700 Fribourg, Switzerland}
\affiliation{$^2$ Physical Chemistry, Department of Chemistry, Lund University, 221 00 Lund, Sweden}

\date{\today}

\begin{abstract}
Hierarchical polymer structures such as pNIPAM microgels have been extensively studied for their ability to undergo significant structural and physical transformations that can be controlled by external stimuli such as temperature, pH or solvent composition. However, direct three-dimensional visualization of individual particles in situ have so far been hindered by insufficient resolution, with optical microscopy, or contrast, with electron microscopy. In recent years superresolution microscopy techniques have emerged that in principle can provide nanoscopic optical resolution. Here we report on the in-situ superresolution microscopy of dye-labeled submicron sized pNIPAM microgels revealing the internal microstructure during swelling and collapse of individual particles.  Using direct STochastic Optical Reconstruction Microscopy (dSTORM) we demonstrate a lateral optical resolution of 30nm and an axial resolution of 60nm.
\end{abstract}

\maketitle


\section{Introduction}
Polymer microgels are a hybrid between a colloid and a polymer and the combination of these two classes of materials offers a number of advantages \cite{fernandez2011microgel}. The well-defined shape and size of a colloid provides control over the microstructural length scales and response times while the polymeric nature offers physico-chemical control parameters that can be sensitive to external stimuli.  The most widely studied material of this kind is based on the polymer Poly(N-isopropylacrylamide) (pNIPAM) \cite{fernandez2011microgel}. It can be readily cross-linked during synthesis to obtain microgel particles with a size that can be controlled in the range 100-1000nm. The vast interest stems from the fact that the polymer is thermosensitive with a lower-critical solution temperature of approximately 32$^{\circ}$C, which is close to physiological conditions \cite{wu1998globule}. The volume phase transition of the microgels can also be induced by addition of alcohols \cite{winnik1990methanol,costa2002phase} and by changes in pH which offers a plethora of possibilities for the design of stimuli responsive materials and for sensing and substance release applications \cite{sanchez2009synthesis,suspensions2011fundamentals,snowden1996colloidal,ballauff2007smart,berndt2003doubly,islam2014poly,contreras2011multifunctional}.
 It has also been argued that the polymer collapse is reminiscent of protein denaturation \cite{dobson1996protein,graziano2000temperature,chen2005folding}. Despite the overwhelming interest a direct visualization of volume phase transition of individual particles in-situ has so far been lacking. \newline The size of microgel particles is typically a micrometer or less and therefore conventional light microscopy cannot resolve its internal structure. Measurements by atomic force microscopy have been reported in the collapsed dry state only \cite{schmidt2008thermoresponsive}. Equally, transmission electron microscopy (TEM) is normally only applied to dried and collapsed microgels \cite{dagallier2010thermoresponsive}.  In one study cryo-TEM has been applied to swollen polystyrene-pNIPAM core-shell particles at a single temperature \cite{crassous2009quantitative}.  This method, although cumbersome and suffering from low contrast, might have potential for the characterization of the internal microstructure of  pNIPAM microgels. Equally modern synchroton based x-ray nanotomography, with a resolution in the 50 nm range, could in principle be used \cite{withers2007x}. Scattering methods using X-ray, neutrons and light do have the required resolving power and have thus been employed frequently \cite{reufer2009temperature,stieger2004small,ballauff2007smart}. However they only provide information about radially averaged properties of an ensemble of particles and thus information on a single part level is not accessible.
\newline \indent In recent years superresolution microscopy techniques have emerged that in principle can provide nanoscopic optical resolution\cite{huang2008three,huang2009super,hell2007far,galbraith2011super,betzig2006imaging,hell2009microscopy,heilemann2008subdiffraction,van2011direct}. Despite their overwhelming popularity in the field of bioimaging very few successful examples for their application in materials sciences are known \cite{aoki2012conformational,berro2012super}. Here we report on the in-situ superresolution microscopy of pNIPAM microgels revealing the internal microstructure during swelling and collapse of individual particles induced by addition of controlled amounts of methanol to the solvent \cite{reufer2009temperature}.

\section{Experimental}
\subsection{Microgel synthesis and labeling}
The microgels are synthesized by free radical precipitation polymerisation in a batch reactor using N-isopropylacrylamide (Acros Organics, 99\%), NIPAM, as the monomeric unit and N-(3-aminopropyl)methacrylamide hydrochloride (Polysciences), APMA, as a co-monomer. The co-monomer is used in order to incorporates free amine groups into the microgels that are further used as conjugation points for the fluorescent biomarker Alexa Fluor 647. N,N'-Methylenebis(acrylamide) (Sigma-Aldrich, 99\%), BIS, is used as a cross-linker. NIPAM is re-crystalized in hexane and all other chemicals are used as received. NIPAM (1.460g) and BIS (0.103g) are dissolved in 85g of H$_2$O. APMA (0.0066g) is dissolved in 10g of H$_2$O. 5g of the APMA solution is added to the reactor and the reaction mixture is left to de-gas under an argon atmosphere for 40 minutes before the temperature is raised to 70 $^{\circ}$C. The initiator, 2,2'-Azobis(2-methylpropionamidine) dihydrochloride (Sigma-Aldrich, 97\%, 0.0365g) is dissolved in 5g of H$_2$O prior to addition to the reaction mixture. 3 minutes after the initiator is added the injection pump is started, injecting the remaining 5g of APMA solution at an addition rate of 0.5ml/min. The synthesis is carried out for 4 h before the reaction mixture is left to cool down over night under constant stirring. The microgels are then dialysed (dialysis tube, MWCO 10.000, Spectrum laboratories) against de-ionized water for two weeks in order to remove unreacted species. The resulting microgels have a degree of cross-linking of 4.9mol\% and a co-monomer concentration of 0.27mol\% assuming complete consumption of all components.
To label the microgels we mix them with an excess amount of fluorescent dye in water inside a clean eppendorf. We then place it on a slowly oscillating platform at room temperature for 2 hours. Multiple cycles of centrifugation and resuspension are then performed to remove unreacted dye.
\subsection{Sample preparation for dSTORM}
Typical experiments last several minutes thus requiring immobilization of the sample. To achieve this we irreversibly adsorb the particles by placing drop of microgel suspension between two glass coverslips, spreading it into a thin layer, then placing it in the oven at 55$^{\circ}$C until they are dry.  The coverslips are previously treated with piranha solution (a mixture of sulfuric acid (H$_2$SO$_4$) and hydrogen peroxide (H$_2$O$_2$)) to clean them thoroughly. The particles are then resuspended in the appropriate water-methanol solution and found to be immobile for the duration of the experiments. The solution also contains Cysteamine (Sigma Aldrich) at 50 mM concentration and the pH is adjusted to 8 using HCl.  These steps are essential to achieve blinking of fluorophores \cite{van2011direct}.
In principle, to achieve optimal blinking and therefore the highest possible resolution one should also add oxygen scavengers such as GLOX \cite{dempsey2011evaluation,van2011direct} to the imaging solution. We decided to only include cysteamine in order to keep the systems as pure as possible, preserving the microgel deswelling behaviour at the expense of fluorophore brightness and photostability.
\subsection{Static and dynamic light scattering}
We use a commercial light scattering goniometer (ALV, Germany) for the characterization of the size of the microgel particles in suspension. A green laser wavelength  ($\lambda$=532nm) is selected because it does not excite the fluorophores present in the microgel structure. We extract the dynamic light scattering hydrodynamic radius $R_h$ from a standard first cumulant fit of the intensity correlation function at three different scattering angles $40,50$ and $60^\circ$. The viscosity of the mixture has been taken from published values \cite{lide2004crc}. The radius and polydispersity measured by means of static light scattering are obtained by fitting the scattered intensity as a function of the scattering vector $q$ to the fuzzy sphere modelof Stieger et al. \cite{stieger2004small}. The contribution of back-reflected light at high scattering angles is taken into account with an additional adjustable parameter as described in ref. \cite{zemb2002neutrons}. To avoid aggregation of the microgels at higher temperatures we lower the Cysteamine content to 20mM  until the suspension is found stable. To verify that Cysteamine has no influence on particle size we also perform measurements at different concentrations, before the onset of aggregation, and find the size to be unchanged.
\subsection{dSTORM Superresolution microscopy}
We apply direct Stochastic Optical Reconstruction Microscopy (dSTORM) for supper resolution imaging as described in \cite{van2011direct}. To this end we use a  897Nikon TiEclipse inverted microscope, equipped with an EMCCD camera (Andor ixon Ultra) and a total internal reflection fluorescence (TIRF) arm to achieve highly inclined illumination situation with limited fluorescence background noise. The illumination is provided by a powerful red laser (Coherent Genesis 1W at 639nm) and a weaker violet one (Toptica 120mW at 405nm), both coupled into a single mode fiber into the TIRF arm. The light is focused on the back aperture of a high numerical aperture and magnification objective (NA 1.49 and 100x magnification), collimating the beam. Illumination with the red laser is used to achieve sparse fluorophore blinking while the violet laser is used in order to tune the density of blinking molecules. An extra zoom lens is placed before the camera to achieve a final pixel size corresponding to an edge length of 104nm at the given magnification. A dichroic filter with a central wavelength of 700nm and bandwidth of 75nm is placed in the detection pathway (ET700/75, Chroma).
\newline \indent For every image we reconstruct we acquire 60 000 frames with 8-10ms exposure time and EM gain set to 300.  The thousands of images are then analyzed using the open source software ThunderSTORM which determines, for every fluorescent spot, the position, brightness and localization precision \cite{ovesny2014thunderstorm}. We keep only points that are localized with a precision below 15nm. To correct for residual small drifts the image stack is subdivided into 20 subsets, each is used to reconstruct an image, then those images are cross correlated to determine drift information. The same software is also used to reconstruct superresolution images \cite{ovesny2014thunderstorm}.
Three-dimensional imaging is performed using the method of astigmatism \cite{huang2008three} which we implement using the adaptive optics microscope add-on MicAO (Imagine Optic, France). With the MicAO in place the point spread function changes shape depending on the axial position of the fluorescent emitter. We calibrate for this by imaging fluorescent beads (100nm diameter Tetraspeck, Termofisher) scanned through the focal plane. Due to the refractive index mismatch between the sample and the glass a focal shift is present and must be corrected for. As shown in \cite{bratton2015simple}, since the sample is within 1.5 microns of the coverslip surface, it is sufficient to rescale all axial positions by a factor of 0.57 when using the same optics.
We determine the nanoscopic resolution from the distribution of measured localizations obtained by the same fluorophore. Although the fluorophore position is immobile it still gives stochastically spread localizations which can be modelled by a Gaussian. The full width half maximum (FWHM) then gives us the lateral resolution which we determine to be 30nm on average. We note that fitting $\rho_{2D}$ for the collapsed microgels with a box profile of constant density equally yields a resolution of $\simeq$30nm. The axial resolution is estimated to about 60nm but this value does not enter our analysis and thus has not been determined with the same accuracy.
\section{Results and discussion}
\begin{figure}  \centering\includegraphics[trim=0cm 0cm 0cm -0.5cm, clip=true, width=8.2cm]{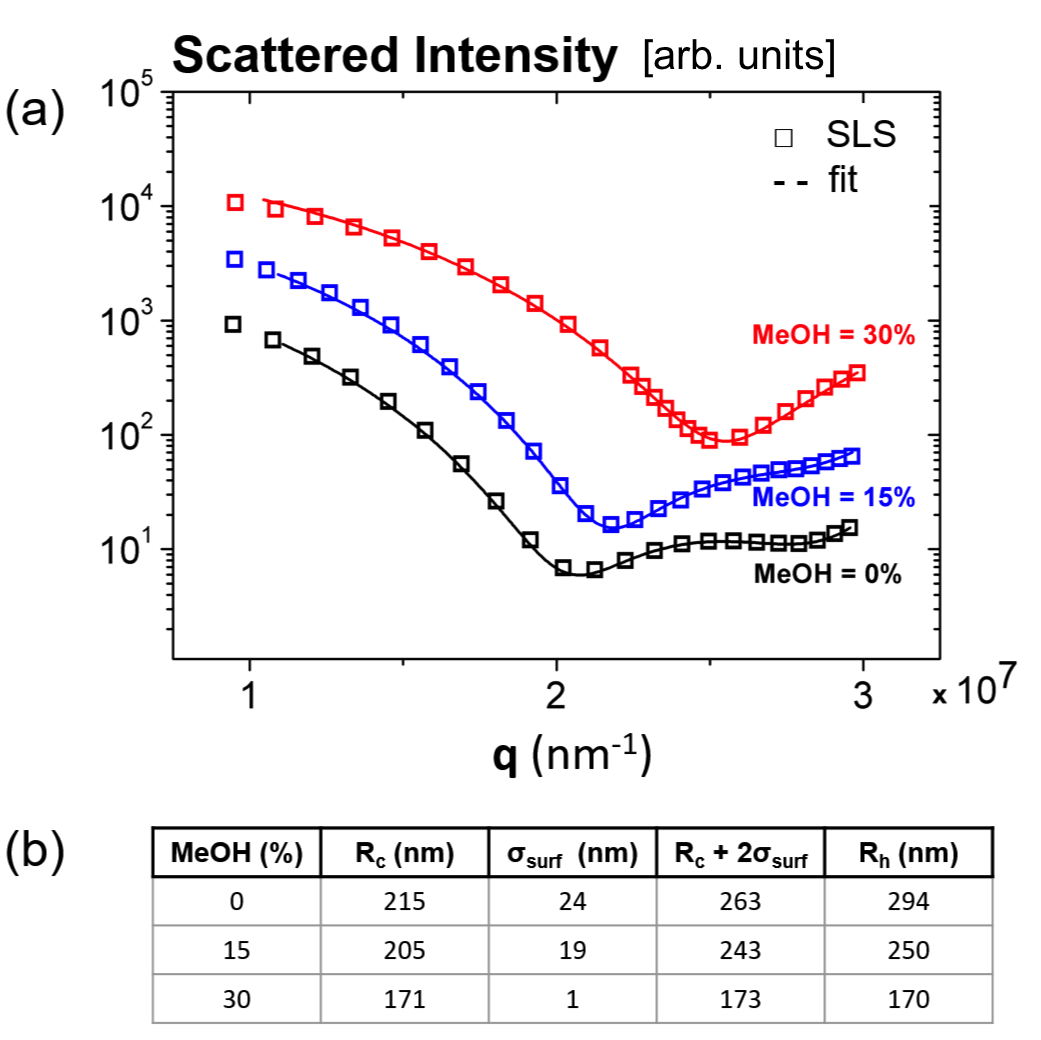}
  \caption{Static and dynamic light scattering characterization of dilute pNIPAM microgel suspensions. (a) Symbols: Static light scattering curves for different methanol contents. Curves are shifted for clarity. Dashed lines: Fit with the fuzzy sphere model for a size polydispersity of $7\%$. (b) Table of results obtained for the hydrodynamic radius $R_h$ from DLS and $R,\sigma_{surf}$ extracted from the fit of the SLS data shown in panel (a).}
  \label{fgr:SLS}
\end{figure}
\subsection{Light scattering characterization of microgel particles} We prepare dilute pNIPAM microgel particle suspensions at room temperature (T=22$^{\circ}$C) in a mixture of water and methanol. Pure water is a good solvent for Poly(N-isopropylacrylamide) at the selected temperature and the microgels are highly swollen. Methanol is chosen as a deswelling agent because it offers a convenient handle to induce the volume phase transition by changing the solvent composition. Moreover, deswelling with methanol does not require setting accurately different temperatures in the microscope sample chamber.  Adding some percentile of methanol induces deswelling until the particles are maximally collapsed for a methanol content of $30\%$ \cite{winnik1990methanol,costa2002phase,bischofberger2014hydrophobic}. We first apply static (SLS) and dynamic light scattering (DLS) to characterize the microgel particles in situ for different solvent compositions. As we add methanol we observe a shift of the minimum of the scattering curve $I(q)$ towards larger values of $q$ and an increase of the scattered intensity, Fig. \ref{fgr:SLS}. These observations show that the microgel particles deswell. We can model the scattering curves quantitatively in the weak scattering or Rayleigh-Gans-Debye approximation considering an isotropic density distribution. To this end we fit the experimental data with the 'fuzzy sphere' model of Stieger et al. \cite{stieger2004small} taking into account polydispersity and internal reflections in the light scattering cuvette. The model is derived by convoluting the box profile of a homogeneous sphere, radius R, with a Gaussian, standard deviation $\sigma _{surf}$. In Fourier space the convolution is represented by a product and the intensity distribution takes the simple form $I\left( q \right) \propto {\left[ {3{{\left[ {\sin \left( {qR} \right) - qR\cos \left( {qR} \right)} \right]} \mathord{\left/ {\vphantom {{\left[ {\sin \left( {qR} \right) - qR\cos \left( {qR} \right)} \right]} {\exp \left( { - {{{{\left( {{\sigma _{surf}}q} \right)}^2}} \mathord{\left/
 {\vphantom {{{{\left( {{\sigma _{surf}}q} \right)}^2}} 2}} \right.
 \kern-\nulldelimiterspace} 2}} \right)}}} \right.
 \kern-\nulldelimiterspace} {\exp \left( { - {{{{\left( {{\sigma _{surf}}q} \right)}^2}} \mathord{\left/
 {\vphantom {{{{\left( {{\sigma _{surf}}q} \right)}^2}} 2}} \right.
 \kern-\nulldelimiterspace} 2}} \right)}}} \right]^2}$. The simplicity of this expression is to a large extent responsible for the success of this model although it has been recognized early-on that the model is not entirely satisfactory as it does not predict a decay to zero density at a finite distance \cite{fernandez2011microgel}. Alternative models for the density distribution have been suggested such as an antisymmetric parabolic shell \cite{berndt2005structure}, a linear shell profile \cite{mason2005density} or a dense core covered by a brush \cite{scheffold2010brushlike,romeo2013elasticity}. The latter describes well the onset of particle-particle interactions but until now has only been discussed for a brush with constant density \cite{scheffold2010brushlike,kim2009compression}.
\newline In practice we find the fit of the fuzzy sphere model to the light scattering data excellent as shown in Fig. \ref{fgr:SLS}. We observe that the particle core radius $R$ shrinks from 215nm to 171nm. At the same time the shell of about $2\sigma_{surf}\sim 50$nm thickness collapses to nearly zero. From dynamic light scattering (DLS) on the same samples we obtain the hydrodynamic radius of the particles $R_h$. As suggested in previous studies we find that $R+2\sigma_{surf} \approx R_h$\cite{stieger2004small}.
\subsection{dSTORM imaging of microgel particles}
\begin{figure}  \centering\includegraphics[width=7cm]{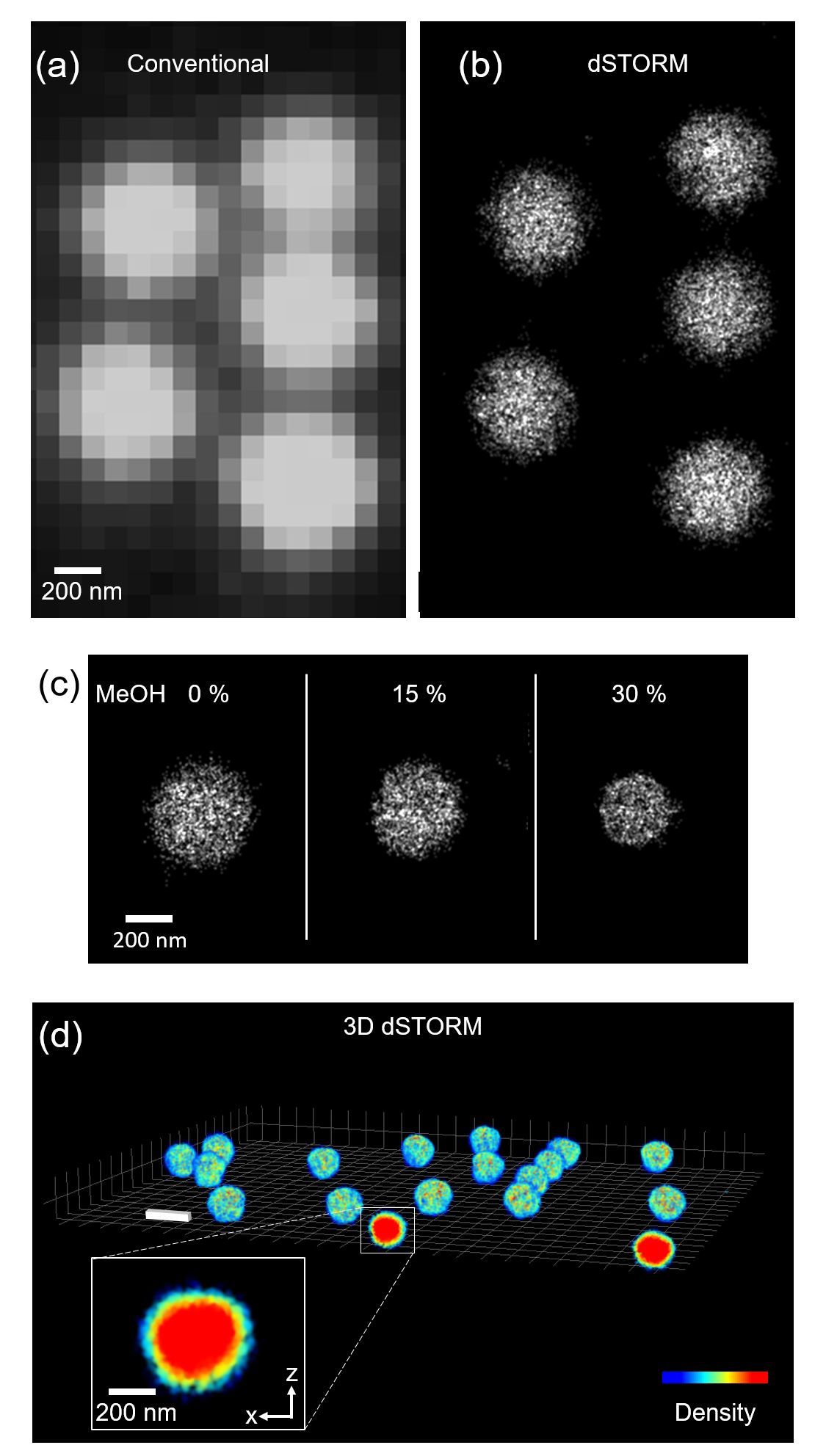}
  \caption{dSTORM superresolution microscopy of pNIPAM microgels at T=$22^\circ$C. (a) Standard widefield fluorescence microscopy image of the swollen microgel particles. (b) dSTORM image of the same sample. (c) dSTORM of the deswelling of the microgel particles upon addition of methanol. (d) 3D image of the swollen microgel particles without the addition of  methanol. Grid size is 300nm, scale bar 600nm. The inset shows an enlarged view of the density distribution inside an individual particle. dSTORM lateral resolution is $\sim$30nm and axial resolution is $\sim$60nm.}
  \label{fgr:dSTORM}
\end{figure}

For the dSTORM characterization the microgel particles need to remain at a fixed position for several minutes. We thus chose to irreversibly adsorb the particles on a microscope slide. To this end we first dry, on a glass coverslip, a small amount of the microgel suspension at 55$^{\circ}$. After resuspension in the appropriate solvent the particles are found to remain immobile.
Next we perform dSTORM \cite{van2011direct} measurements by homogeneously illuminating the sample at 639nm under nearly total internal reflection conditions, Fig. \ref{fgr:dSTORM}.  For comparison we also show an image taken with conventional wide-field microscopy under the same conditions. Additional illumination at 405nm is added when necessary to maintain a sufficient density of blinking fluorophores. For each image we acquire 60'000 frames with an 8-10ms exposure, resulting in measurement times of 8-10 minutes.
Image analysis and reconstruction is performed using the freely available software ThunderSTORM \cite{ovesny2014thunderstorm}. For each frame, images of single fluorophores are fit to 2D gaussians, determining the x-y position, intensity and localization precision. We remove points which are localized with a precision less than 15nm, and then use the remaining ones to reconstruct superresolution images as shown in Fig.  \ref{fgr:dSTORM}. Drift is corrected using the method of cross-correlation. From repeated localizations of the same fluorophores we estimate a resolution of 30nm in the plane. We obtain 3000 - 5000 fluorophore localizations per particle with typically one to two thousand photons detected per localization.
Several methods are available to obtain three-dimensional superresolution images \cite{huang2008three, juette2008three, pavani2009three, deschamps20143d, shtengel2009interferometric}, out of those we chose that of astigmatism, due to its relative simplicity and availability of software for analysis. With this method the point spread function is distorted in a controlled way, encoding axial position information into its shape. Previous calibration of the distortion allows for the reconstruction of three-dimensional images, as shown in Fig.  \ref{fgr:dSTORM}. With this method the axial resolution is lower than in the plane and is estimated to be 60nm, still an order of magnitude better than confocal microscopy.
Such 3D imaging on the nanoscale might not be crucial for more or less isotropic microgels but will become of key importance whenever the particle architecture is more complex such as for ellipsoidal particles or heterogeneous structures \cite{karg2007nanorod,dechezelles2013hybrid,crassous2015anisotropic}. In our experiments we do not observe a strong anisotropy of the particles and we therefore decided to take advantage of the higher lateral resolution for the quantitative analysis of the radial density distribution inside the microgel particle. All density profiles are thus modelled based on the 2D projection of detected sites on the plane of observation.
\begin{figure}  \centering\includegraphics[trim=0cm 0.5cm 0cm 0cm, clip=true, width=8.2cm]{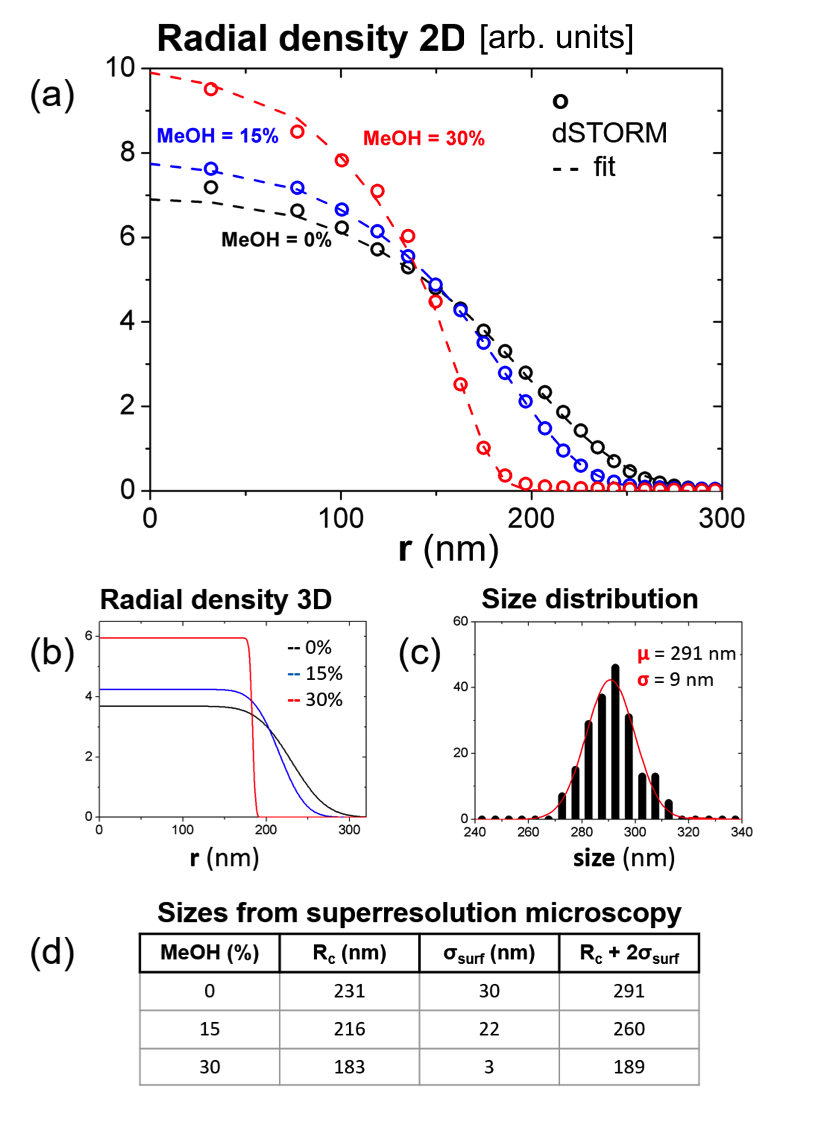}
  \caption{dSTORM analysis of the microgel internal structure. (a) Symbols: measured 2D density profiles (average over about one hundred particles). Dashed lines: fit with the fuzzy sphere model with a core radius $R$ and a shell parameter $\sigma _{surf}$. (b) 3D radial density profiles corresponding to the fits shown in (a). (c) Distribution of size $R+2\sigma _{surf}$ obtained from fitting the density profile of individual particles. Average size $\mu=$$291$nm, standard deviation $\sigma=9$nm. (d) Table of results for the mean particle core radius and shell determined by dSTORM.}
  \label{fgr:radprofile}
\end{figure}
\subsection{Microgel density profiles}
In Fig. \ref{fgr:radprofile} we show the measured ensemble averaged 2D density profiles $\rho_{2D}(r)$ for three different solvent compositions. Note that due to the projection a homogeneous sphere with radius $R$ appears as $\rho_{2D}(r)\propto \sqrt(R^2-r^2);r\le R$. Moreover we have to include the finite resolution $\xi\simeq 30$nm in our analysis.  Next we fit images of the swollen microgels while keeping $\xi$ constant. We apply the fuzzy sphere model by Stieger at al. \cite{stieger2004small} that we had used to model the static light scattering curves, Fig. \ref{fgr:SLS}. The adjustable fit parameters are the core radius $R$ and the smearing parameter $\sigma_{surf}$ which corresponds to about half the thickness of the fuzzy shell. In 3D the model predicts a density distribution $\rho(r)\propto erfc(r-R,\sigma_{surf})$. For the fit we convolute this function to take account for the finite resolution $\xi=30$nm, project it to the plane and adjust $R,\sigma_{surf}$ until we obtain a best fit (see also supporting information). The results are shown in Fig.   \ref{fgr:radprofile} as dashed lines. The core radius and the smearing parameter $\sigma_{surf}$ extracted from the fit are in quantitative agreement with the scattering results. The corresponding radial density profiles in 3D are shown in the inset of Fig. \ref{fgr:radprofile}.
In Fig. \ref{fgr:radprofile} c) we show the results for the particle size obtained by analysing the density profiles of individual microgels. From this data we can extract the polydispersity (standard deviation divided by the mean) of the particle size $\sim R + 2\sigma_{surf}$ and find a value of $3.1\%$. This value is somewhat smaller than the one obtained from static light scattering, Figure \ref{fgr:SLS}. We believe this discrepancy is due to the failure of the Rayleigh Gans Debye approximation in light scattering due to the finite refractive index contrast even for the low density swollen microgels \cite{reufer2009temperature}. The latter result is a hint for the power of a visualization of the individual microgels: It allows a direct real-space interpretation of the experimental results and thus in many cases can provide more accurate information.

\section{Summary and Conclusions}
In summary we could demonstrate the successful application of dSTORM superresolution microscopy to stimuli-repsonsive pNIPAM microgels. For the microgels studied the dSTORM results are in quantitative agreement with static and dynamic light scattering. Our results can serve as a benchmark and open the pathway for future applications towards more complex polymer and microgel architectures. Anisotropic or core-shell microgels as well as microgels that are doped or decorated with metal nanoparticles for example are difficult or impossible to characterize in-situ with conventional techniques \cite{dechezelles2013hybrid}. A generalization of our approach using multicolor dSTORM \cite{lampe2012multi} however can be applied in a next step for a detailed characterization and co-localization of functional, chemically distinct subunits on the nanoscale.

\section{Acknowledgements}
This work was supported by the Swiss National Science Foundation through project number 149867, and through the National Centre of Competence in Research \emph{Bio-Inspired Materials}. We acknowledge financial support by the Adolphe Merkle foundation. We thank Isabelle Sp\"uhler, Georges Br\"ugger, Veronique Trappe and Davide Calzolari for fruitful discussions. PS gratefully acknowledges support from the European Research Council (ERC-339678-COMPASS) and the Swedish Research Council (621-2014-4037).



\bibliographystyle{rsc} 

\begin{thebibliography}{50}
\providecommand*{\natexlab}[1]{#1}
\providecommand*{\mciteSetBstSublistMode}[1]{}
\providecommand*{\mciteSetBstMaxWidthForm}[2]{}
\providecommand*{\mciteBstWouldAddEndPuncttrue}
  {\def\EndOfBibitem{\unskip.}}
\providecommand*{\mciteBstWouldAddEndPunctfalse}
  {\let\EndOfBibitem\relax}
\providecommand*{\mciteSetBstMidEndSepPunct}[3]{}
\providecommand*{\mciteSetBstSublistLabelBeginEnd}[3]{}
\providecommand*{\EndOfBibitem}{}
\mciteSetBstSublistMode{f}
\mciteSetBstMaxWidthForm{subitem}
{(\emph{\alph{mcitesubitemcount}})}
\mciteSetBstSublistLabelBeginEnd{\mcitemaxwidthsubitemform\space}
{\relax}{\relax}
\bibitem[Fernandez-Nieves \emph{et~al.}(2011)Fernandez-Nieves, Wyss, Mattsson,
  and Weitz]{fernandez2011microgel}
A.~Fernandez-Nieves, H.~Wyss, J.~Mattsson and D.~A. Weitz, \emph{Microgel
  suspensions: fundamentals and applications}, John Wiley \& Sons, 2011\relax
\mciteBstWouldAddEndPuncttrue
\mciteSetBstMidEndSepPunct{\mcitedefaultmidpunct}
{\mcitedefaultendpunct}{\mcitedefaultseppunct}\relax
\EndOfBibitem
\bibitem[Wu and Wang(1998)]{wu1998globule}
C.~Wu and X.~Wang, \emph{Physical Review Letters}, 1998, \textbf{80},
  4092\relax
\mciteBstWouldAddEndPuncttrue
\mciteSetBstMidEndSepPunct{\mcitedefaultmidpunct}
{\mcitedefaultendpunct}{\mcitedefaultseppunct}\relax
\EndOfBibitem
\bibitem[Winnik \emph{et~al.}(1990)Winnik, Ringsdorf, and
  Venzmer]{winnik1990methanol}
F.~M. Winnik, H.~Ringsdorf and J.~Venzmer, \emph{Macromolecules}, 1990,
  \textbf{23}, 2415--2416\relax
\mciteBstWouldAddEndPuncttrue
\mciteSetBstMidEndSepPunct{\mcitedefaultmidpunct}
{\mcitedefaultendpunct}{\mcitedefaultseppunct}\relax
\EndOfBibitem
\bibitem[Costa and Freitas(2002)]{costa2002phase}
R.~O. Costa and R.~F. Freitas, \emph{Polymer}, 2002, \textbf{43},
  5879--5885\relax
\mciteBstWouldAddEndPuncttrue
\mciteSetBstMidEndSepPunct{\mcitedefaultmidpunct}
{\mcitedefaultendpunct}{\mcitedefaultseppunct}\relax
\EndOfBibitem
\bibitem[sus(2011)]{suspensions2011fundamentals}
\emph{Microgel Suspensions: Fundamentals and Applications, edited by A.
  Fernandez de Las Nieves, H. Wyss, J. Mattson, and DA Weitz}, 2011\relax
\mciteBstWouldAddEndPuncttrue
\mciteSetBstMidEndSepPunct{\mcitedefaultmidpunct}
{\mcitedefaultendpunct}{\mcitedefaultseppunct}\relax
\EndOfBibitem
\bibitem[Snowden \emph{et~al.}(1996)Snowden, Chowdhry, Vincent, and
  Morris]{snowden1996colloidal}
M.~J. Snowden, B.~Z. Chowdhry, B.~Vincent and G.~E. Morris, \emph{Journal of
  the Chemical Society, Faraday Transactions}, 1996, \textbf{92},
  5013--5016\relax
\mciteBstWouldAddEndPuncttrue
\mciteSetBstMidEndSepPunct{\mcitedefaultmidpunct}
{\mcitedefaultendpunct}{\mcitedefaultseppunct}\relax
\EndOfBibitem
\bibitem[Ballauff and Lu(2007)]{ballauff2007smart}
M.~Ballauff and Y.~Lu, \emph{Polymer}, 2007, \textbf{48}, 1815--1823\relax
\mciteBstWouldAddEndPuncttrue
\mciteSetBstMidEndSepPunct{\mcitedefaultmidpunct}
{\mcitedefaultendpunct}{\mcitedefaultseppunct}\relax
\EndOfBibitem
\bibitem[Berndt and Richtering(2003)]{berndt2003doubly}
I.~Berndt and W.~Richtering, \emph{Macromolecules}, 2003, \textbf{36},
  8780--8785\relax
\mciteBstWouldAddEndPuncttrue
\mciteSetBstMidEndSepPunct{\mcitedefaultmidpunct}
{\mcitedefaultendpunct}{\mcitedefaultseppunct}\relax
\EndOfBibitem
\bibitem[Islam \emph{et~al.}(2014)Islam, Ahiabu, Li, and Serpe]{islam2014poly}
M.~R. Islam, A.~Ahiabu, X.~Li and M.~J. Serpe, \emph{Sensors}, 2014,
  \textbf{14}, 8984--8995\relax
\mciteBstWouldAddEndPuncttrue
\mciteSetBstMidEndSepPunct{\mcitedefaultmidpunct}
{\mcitedefaultendpunct}{\mcitedefaultseppunct}\relax
\EndOfBibitem
\bibitem{sanchez2009synthesis} A. S{\'a}nchez-Iglesias, et al., {\it{ACS nano}} {\bf 3}(10), 3184-3190 (2009).
\bibitem[Contreras-C{\'a}ceres \emph{et~al.}(2011)Contreras-C{\'a}ceres,
  Abalde-Cela, Guardia-Gir{\'o}s, Fern{\'a}ndez-Barbero, P{\'e}rez-Juste,
  Alvarez-Puebla, and Liz-Marz{\'a}n]{contreras2011multifunctional}
R.~Contreras-C{\'a}ceres, S.~Abalde-Cela, P.~Guardia-Gir{\'o}s,
  A.~Fern{\'a}ndez-Barbero, J.~P{\'e}rez-Juste, R.~A. Alvarez-Puebla and L.~M.
  Liz-Marz{\'a}n, \emph{Langmuir}, 2011, \textbf{27}, 4520--4525\relax
\mciteBstWouldAddEndPuncttrue
\mciteSetBstMidEndSepPunct{\mcitedefaultmidpunct}
{\mcitedefaultendpunct}{\mcitedefaultseppunct}\relax
\EndOfBibitem
\bibitem[Nolan \emph{et~al.}(2004)Nolan, Serpe, and Lyon]{nolan2004thermally}
C.~M. Nolan, M.~J. Serpe and L.~A. Lyon, \emph{Biomacromolecules}, 2004,
  \textbf{5}, 1940--1946\relax
\mciteBstWouldAddEndPuncttrue
\mciteSetBstMidEndSepPunct{\mcitedefaultmidpunct}
{\mcitedefaultendpunct}{\mcitedefaultseppunct}\relax
\EndOfBibitem
\bibitem[Dobson and Fersht(1996)]{dobson1996protein}
C.~M. Dobson and A.~Fersht, \emph{Protein folding}, Cambridge Univ Pr,
  1996\relax
\mciteBstWouldAddEndPuncttrue
\mciteSetBstMidEndSepPunct{\mcitedefaultmidpunct}
{\mcitedefaultendpunct}{\mcitedefaultseppunct}\relax
\EndOfBibitem
\bibitem[Graziano(2000)]{graziano2000temperature}
G.~Graziano, \emph{International journal of biological macromolecules}, 2000,
  \textbf{27}, 89--97\relax
\mciteBstWouldAddEndPuncttrue
\mciteSetBstMidEndSepPunct{\mcitedefaultmidpunct}
{\mcitedefaultendpunct}{\mcitedefaultseppunct}\relax
\EndOfBibitem
\bibitem[Chen \emph{et~al.}(2005)Chen, Li, Ding, Zhang, Zhang, and
  Wu]{chen2005folding}
H.~Chen, J.~Li, Y.~Ding, G.~Zhang, Q.~Zhang and C.~Wu, \emph{Macromolecules},
  2005, \textbf{38}, 4403--4408\relax
\mciteBstWouldAddEndPuncttrue
\mciteSetBstMidEndSepPunct{\mcitedefaultmidpunct}
{\mcitedefaultendpunct}{\mcitedefaultseppunct}\relax
\EndOfBibitem
\bibitem[Schmidt \emph{et~al.}(2008)Schmidt, Motschmann, Hellweg, and von
  Klitzing]{schmidt2008thermoresponsive}
S.~Schmidt, H.~Motschmann, T.~Hellweg and R.~von Klitzing, \emph{Polymer},
  2008, \textbf{49}, 749--756\relax
\mciteBstWouldAddEndPuncttrue
\mciteSetBstMidEndSepPunct{\mcitedefaultmidpunct}
{\mcitedefaultendpunct}{\mcitedefaultseppunct}\relax
\EndOfBibitem
\bibitem[Dagallier \emph{et~al.}(2010)Dagallier, Dietsch, Schurtenberger, and
  Scheffold]{dagallier2010thermoresponsive}
C.~Dagallier, H.~Dietsch, P.~Schurtenberger and F.~Scheffold, \emph{Soft
  Matter}, 2010, \textbf{6}, 2174--2177\relax
\mciteBstWouldAddEndPuncttrue
\mciteSetBstMidEndSepPunct{\mcitedefaultmidpunct}
{\mcitedefaultendpunct}{\mcitedefaultseppunct}\relax
\EndOfBibitem
\bibitem[Crassous \emph{et~al.}(2009)Crassous, Rochette, Wittemann, Schrinner,
  Ballauff, and Drechsler]{crassous2009quantitative}
J.~J. Crassous, C.~N. Rochette, A.~Wittemann, M.~Schrinner, M.~Ballauff and
  M.~Drechsler, \emph{Langmuir}, 2009, \textbf{25}, 7862--7871\relax
\mciteBstWouldAddEndPuncttrue
\mciteSetBstMidEndSepPunct{\mcitedefaultmidpunct}
{\mcitedefaultendpunct}{\mcitedefaultseppunct}\relax
\EndOfBibitem
\bibitem[Withers(2007)]{withers2007x}
P.~J. Withers, \emph{Materials today}, 2007, \textbf{10}, 26--34\relax
\mciteBstWouldAddEndPuncttrue
\mciteSetBstMidEndSepPunct{\mcitedefaultmidpunct}
{\mcitedefaultendpunct}{\mcitedefaultseppunct}\relax
\EndOfBibitem
\bibitem[Reufer \emph{et~al.}(2009)Reufer, D{\i}az-Leyva, Lynch, and
  Scheffold]{reufer2009temperature}
M.~Reufer, P.~D{\i}az-Leyva, I.~Lynch and F.~Scheffold, \emph{The European
  Physical Journal E}, 2009, \textbf{28}, 165--171\relax
\mciteBstWouldAddEndPuncttrue
\mciteSetBstMidEndSepPunct{\mcitedefaultmidpunct}
{\mcitedefaultendpunct}{\mcitedefaultseppunct}\relax
\EndOfBibitem
\bibitem[Stieger \emph{et~al.}(2004)Stieger, Richtering, Pedersen, and
  Lindner]{stieger2004small}
M.~Stieger, W.~Richtering, J.~S. Pedersen and P.~Lindner, \emph{The Journal of
  chemical physics}, 2004, \textbf{120}, 6197--6206\relax
\mciteBstWouldAddEndPuncttrue
\mciteSetBstMidEndSepPunct{\mcitedefaultmidpunct}
{\mcitedefaultendpunct}{\mcitedefaultseppunct}\relax
\EndOfBibitem
\bibitem[Huang \emph{et~al.}(2008)Huang, Wang, Bates, and
  Zhuang]{huang2008three}
B.~Huang, W.~Wang, M.~Bates and X.~Zhuang, \emph{Science}, 2008, \textbf{319},
  810--813\relax
\mciteBstWouldAddEndPuncttrue
\mciteSetBstMidEndSepPunct{\mcitedefaultmidpunct}
{\mcitedefaultendpunct}{\mcitedefaultseppunct}\relax
\EndOfBibitem
\bibitem[Huang \emph{et~al.}(2009)Huang, Bates, and Zhuang]{huang2009super}
B.~Huang, M.~Bates and X.~Zhuang, \emph{Annual review of biochemistry}, 2009,
  \textbf{78}, 993\relax
\mciteBstWouldAddEndPuncttrue
\mciteSetBstMidEndSepPunct{\mcitedefaultmidpunct}
{\mcitedefaultendpunct}{\mcitedefaultseppunct}\relax
\EndOfBibitem
\bibitem[Hell(2007)]{hell2007far}
S.~W. Hell, \emph{science}, 2007, \textbf{316}, 1153--1158\relax
\mciteBstWouldAddEndPuncttrue
\mciteSetBstMidEndSepPunct{\mcitedefaultmidpunct}
{\mcitedefaultendpunct}{\mcitedefaultseppunct}\relax
\EndOfBibitem
\bibitem[Galbraith and Galbraith(2011)]{galbraith2011super}
C.~G. Galbraith and J.~A. Galbraith, \emph{Journal of cell science}, 2011,
  \textbf{124}, 1607--1611\relax
\mciteBstWouldAddEndPuncttrue
\mciteSetBstMidEndSepPunct{\mcitedefaultmidpunct}
{\mcitedefaultendpunct}{\mcitedefaultseppunct}\relax
\EndOfBibitem
\bibitem[Betzig \emph{et~al.}(2006)Betzig, Patterson, Sougrat, Lindwasser,
  Olenych, Bonifacino, Davidson, Lippincott-Schwartz, and
  Hess]{betzig2006imaging}
E.~Betzig, G.~H. Patterson, R.~Sougrat, O.~W. Lindwasser, S.~Olenych, J.~S.
  Bonifacino, M.~W. Davidson, J.~Lippincott-Schwartz and H.~F. Hess,
  \emph{Science}, 2006, \textbf{313}, 1642--1645\relax
\mciteBstWouldAddEndPuncttrue
\mciteSetBstMidEndSepPunct{\mcitedefaultmidpunct}
{\mcitedefaultendpunct}{\mcitedefaultseppunct}\relax
\EndOfBibitem
\bibitem[Hell(2009)]{hell2009microscopy}
S.~W. Hell, \emph{Nature methods}, 2009, \textbf{6}, 24--32\relax
\mciteBstWouldAddEndPuncttrue
\mciteSetBstMidEndSepPunct{\mcitedefaultmidpunct}
{\mcitedefaultendpunct}{\mcitedefaultseppunct}\relax
\EndOfBibitem
\bibitem[Heilemann \emph{et~al.}(2008)Heilemann, van~de Linde, Sch{\"u}ttpelz,
  Kasper, Seefeldt, Mukherjee, Tinnefeld, and
  Sauer]{heilemann2008subdiffraction}
M.~Heilemann, S.~van~de Linde, M.~Sch{\"u}ttpelz, R.~Kasper, B.~Seefeldt,
  A.~Mukherjee, P.~Tinnefeld and M.~Sauer, \emph{Angewandte Chemie
  International Edition}, 2008, \textbf{47}, 6172--6176\relax
\mciteBstWouldAddEndPuncttrue
\mciteSetBstMidEndSepPunct{\mcitedefaultmidpunct}
{\mcitedefaultendpunct}{\mcitedefaultseppunct}\relax
\EndOfBibitem
\bibitem[van~de Linde \emph{et~al.}(2011)van~de Linde, L{\"o}schberger, Klein,
  Heidbreder, Wolter, Heilemann, and Sauer]{van2011direct}
S.~van~de Linde, A.~L{\"o}schberger, T.~Klein, M.~Heidbreder, S.~Wolter,
  M.~Heilemann and M.~Sauer, \emph{Nature protocols}, 2011, \textbf{6},
  991--1009\relax
\mciteBstWouldAddEndPuncttrue
\mciteSetBstMidEndSepPunct{\mcitedefaultmidpunct}
{\mcitedefaultendpunct}{\mcitedefaultseppunct}\relax
\EndOfBibitem
\bibitem[Aoki \emph{et~al.}(2012)Aoki, Mori, and Ito]{aoki2012conformational}
H.~Aoki, K.~Mori and S.~Ito, \emph{Soft Matter}, 2012, \textbf{8},
  4390--4395\relax
\mciteBstWouldAddEndPuncttrue
\mciteSetBstMidEndSepPunct{\mcitedefaultmidpunct}
{\mcitedefaultendpunct}{\mcitedefaultseppunct}\relax
\EndOfBibitem
\bibitem[Berro \emph{et~al.}(2012)Berro, Berglund, Carmichael, Kim, and
  Liddle]{berro2012super}
A.~J. Berro, A.~J. Berglund, P.~T. Carmichael, J.~S. Kim and J.~A. Liddle,
  \emph{ACS nano}, 2012, \textbf{6}, 9496--9502\relax
\mciteBstWouldAddEndPuncttrue
\mciteSetBstMidEndSepPunct{\mcitedefaultmidpunct}
{\mcitedefaultendpunct}{\mcitedefaultseppunct}\relax
\EndOfBibitem
\bibitem[Dempsey \emph{et~al.}(2011)Dempsey, Vaughan, Chen, Bates, and
  Zhuang]{dempsey2011evaluation}
G.~T. Dempsey, J.~C. Vaughan, K.~H. Chen, M.~Bates and X.~Zhuang, \emph{Nature
  methods}, 2011, \textbf{8}, 1027--1036\relax
\mciteBstWouldAddEndPuncttrue
\mciteSetBstMidEndSepPunct{\mcitedefaultmidpunct}
{\mcitedefaultendpunct}{\mcitedefaultseppunct}\relax
\EndOfBibitem
\bibitem[Lide(2004)]{lide2004crc}
D.~R. Lide, \emph{CRC handbook of chemistry and physics}, CRC press, 2004\relax
\mciteBstWouldAddEndPuncttrue
\mciteSetBstMidEndSepPunct{\mcitedefaultmidpunct}
{\mcitedefaultendpunct}{\mcitedefaultseppunct}\relax
\EndOfBibitem
\bibitem[Zemb and Lindner(2002)]{zemb2002neutrons}
T.~Zemb and P.~Lindner, \emph{Neutrons, X-rays and light: scattering methods
  applied to soft condensed matter}, North-Holland, 2002\relax
\mciteBstWouldAddEndPuncttrue
\mciteSetBstMidEndSepPunct{\mcitedefaultmidpunct}
{\mcitedefaultendpunct}{\mcitedefaultseppunct}\relax
\EndOfBibitem
\bibitem[Ovesn{\`y} \emph{et~al.}(2014)Ovesn{\`y}, K{\v{r}}{\'\i}{\v{z}}ek,
  Borkovec, {\v{S}}vindrych, and Hagen]{ovesny2014thunderstorm}
M.~Ovesn{\`y}, P.~K{\v{r}}{\'\i}{\v{z}}ek, J.~Borkovec, Z.~{\v{S}}vindrych and
  G.~M. Hagen, \emph{Bioinformatics}, 2014, \textbf{30}, 2389--2390\relax
\mciteBstWouldAddEndPuncttrue
\mciteSetBstMidEndSepPunct{\mcitedefaultmidpunct}
{\mcitedefaultendpunct}{\mcitedefaultseppunct}\relax
\EndOfBibitem
\bibitem[Bratton and Shaevitz(2015)]{bratton2015simple}
B.~P. Bratton and J.~W. Shaevitz, \emph{PloS one}, 2015, \textbf{10},
  e0134616\relax
\mciteBstWouldAddEndPuncttrue
\mciteSetBstMidEndSepPunct{\mcitedefaultmidpunct}
{\mcitedefaultendpunct}{\mcitedefaultseppunct}\relax
\EndOfBibitem
\bibitem[Bischofberger \emph{et~al.}(2014)Bischofberger, Calzolari,
  De~Los~Rios, Jelezarov, and Trappe]{bischofberger2014hydrophobic}
I.~Bischofberger, D.~Calzolari, P.~De~Los~Rios, I.~Jelezarov and V.~Trappe,
  \emph{Scientific reports}, 2014, \textbf{4}, year\relax
\mciteBstWouldAddEndPuncttrue
\mciteSetBstMidEndSepPunct{\mcitedefaultmidpunct}
{\mcitedefaultendpunct}{\mcitedefaultseppunct}\relax
\EndOfBibitem
\bibitem[Berndt \emph{et~al.}(2005)Berndt, Pedersen, and
  Richtering]{berndt2005structure}
I.~Berndt, J.~S. Pedersen and W.~Richtering, \emph{Journal of the American
  Chemical Society}, 2005, \textbf{127}, 9372--9373\relax
\mciteBstWouldAddEndPuncttrue
\mciteSetBstMidEndSepPunct{\mcitedefaultmidpunct}
{\mcitedefaultendpunct}{\mcitedefaultseppunct}\relax
\EndOfBibitem
\bibitem[Mason and Lin(2005)]{mason2005density}
T.~Mason and M.~Lin, \emph{Physical Review E}, 2005, \textbf{71}, 040801\relax
\mciteBstWouldAddEndPuncttrue
\mciteSetBstMidEndSepPunct{\mcitedefaultmidpunct}
{\mcitedefaultendpunct}{\mcitedefaultseppunct}\relax
\EndOfBibitem
\bibitem[Scheffold \emph{et~al.}(2010)Scheffold, D{\'\i}az-Leyva, Reufer,
  Braham, Lynch, and Harden]{scheffold2010brushlike}
F.~Scheffold, P.~D{\'\i}az-Leyva, M.~Reufer, N.~B. Braham, I.~Lynch and J.~L.
  Harden, \emph{Physical Review Letters}, 2010, \textbf{104}, 128304\relax
\mciteBstWouldAddEndPuncttrue
\mciteSetBstMidEndSepPunct{\mcitedefaultmidpunct}
{\mcitedefaultendpunct}{\mcitedefaultseppunct}\relax
\EndOfBibitem
\bibitem[Romeo and Ciamarra(2013)]{romeo2013elasticity}
G.~Romeo and M.~P. Ciamarra, \emph{Soft Matter}, 2013, \textbf{9},
  5401--5406\relax
\mciteBstWouldAddEndPuncttrue
\mciteSetBstMidEndSepPunct{\mcitedefaultmidpunct}
{\mcitedefaultendpunct}{\mcitedefaultseppunct}\relax
\EndOfBibitem
\bibitem[Kim and Matsen(2009)]{kim2009compression}
J.~U. Kim and M.~W. Matsen, \emph{Macromolecules}, 2009, \textbf{42},
  3430--3432\relax
\mciteBstWouldAddEndPuncttrue
\mciteSetBstMidEndSepPunct{\mcitedefaultmidpunct}
{\mcitedefaultendpunct}{\mcitedefaultseppunct}\relax
\EndOfBibitem
\bibitem[Juette \emph{et~al.}(2008)Juette, Gould, Lessard, Mlodzianoski,
  Nagpure, Bennett, Hess, and Bewersdorf]{juette2008three}
M.~F. Juette, T.~J. Gould, M.~D. Lessard, M.~J. Mlodzianoski, B.~S. Nagpure,
  B.~T. Bennett, S.~T. Hess and J.~Bewersdorf, \emph{Nature methods}, 2008,
  \textbf{5}, 527--529\relax
\mciteBstWouldAddEndPuncttrue
\mciteSetBstMidEndSepPunct{\mcitedefaultmidpunct}
{\mcitedefaultendpunct}{\mcitedefaultseppunct}\relax
\EndOfBibitem
\bibitem[Pavani \emph{et~al.}(2009)Pavani, Thompson, Biteen, Lord, Liu, Twieg,
  Piestun, and Moerner]{pavani2009three}
S.~R.~P. Pavani, M.~A. Thompson, J.~S. Biteen, S.~J. Lord, N.~Liu, R.~J. Twieg,
  R.~Piestun and W.~Moerner, \emph{Proceedings of the National Academy of
  Sciences}, 2009, \textbf{106}, 2995--2999\relax
\mciteBstWouldAddEndPuncttrue
\mciteSetBstMidEndSepPunct{\mcitedefaultmidpunct}
{\mcitedefaultendpunct}{\mcitedefaultseppunct}\relax
\EndOfBibitem
\bibitem[Deschamps \emph{et~al.}(2014)Deschamps, Mund, and
  Ries]{deschamps20143d}
J.~Deschamps, M.~Mund and J.~Ries, \emph{Optics Express}, 2014, \textbf{22},
  29081--29091\relax
\mciteBstWouldAddEndPuncttrue
\mciteSetBstMidEndSepPunct{\mcitedefaultmidpunct}
{\mcitedefaultendpunct}{\mcitedefaultseppunct}\relax
\EndOfBibitem
\bibitem[Shtengel \emph{et~al.}(2009)Shtengel, Galbraith, Galbraith,
  Lippincott-Schwartz, Gillette, Manley, Sougrat, Waterman, Kanchanawong,
  Davidson,\emph{et~al.}]{shtengel2009interferometric}
G.~Shtengel, J.~A. Galbraith, C.~G. Galbraith, J.~Lippincott-Schwartz, J.~M.
  Gillette, S.~Manley, R.~Sougrat, C.~M. Waterman, P.~Kanchanawong, M.~W.
  Davidson \emph{et~al.}, \emph{Proceedings of the National Academy of
  Sciences}, 2009, \textbf{106}, 3125--3130\relax
\mciteBstWouldAddEndPuncttrue
\mciteSetBstMidEndSepPunct{\mcitedefaultmidpunct}
{\mcitedefaultendpunct}{\mcitedefaultseppunct}\relax
\EndOfBibitem
\bibitem[Karg \emph{et~al.}(2007)Karg, Pastoriza-Santos, P{\'e}rez-Juste,
  Hellweg, and Liz-Marz{\'a}n]{karg2007nanorod}
M.~Karg, I.~Pastoriza-Santos, J.~P{\'e}rez-Juste, T.~Hellweg and L.~M.
  Liz-Marz{\'a}n, \emph{Small}, 2007, \textbf{3}, 1222--1229\relax
\mciteBstWouldAddEndPuncttrue
\mciteSetBstMidEndSepPunct{\mcitedefaultmidpunct}
{\mcitedefaultendpunct}{\mcitedefaultseppunct}\relax
\EndOfBibitem
\bibitem[Dech{\'e}zelles \emph{et~al.}(2013)Dech{\'e}zelles, Malik, Crassous,
  and Schurtenberger]{dechezelles2013hybrid}
J.-F. Dech{\'e}zelles, V.~Malik, J.~J. Crassous and P.~Schurtenberger,
  \emph{Soft Matter}, 2013, \textbf{9}, 2798--2802\relax
\mciteBstWouldAddEndPuncttrue
\mciteSetBstMidEndSepPunct{\mcitedefaultmidpunct}
{\mcitedefaultendpunct}{\mcitedefaultseppunct}\relax
\EndOfBibitem
\bibitem[Crassous \emph{et~al.}(2015)Crassous, Mihut, M{\aa}nsson, and
  Schurtenberger]{crassous2015anisotropic}
J.~J. Crassous, A.~M. Mihut, L.~K. M{\aa}nsson and P.~Schurtenberger,
  \emph{Nanoscale}, 2015, \textbf{7}, 15971--15982\relax
\mciteBstWouldAddEndPuncttrue
\mciteSetBstMidEndSepPunct{\mcitedefaultmidpunct}
{\mcitedefaultendpunct}{\mcitedefaultseppunct}\relax
\EndOfBibitem
\bibitem[Lampe \emph{et~al.}(2012)Lampe, Haucke, Sigrist, Heilemann, and
  Schmoranzer]{lampe2012multi}
A.~Lampe, V.~Haucke, S.~J. Sigrist, M.~Heilemann and J.~Schmoranzer,
  \emph{Biology of the Cell}, 2012, \textbf{104}, 229--237\relax
\mciteBstWouldAddEndPuncttrue
\mciteSetBstMidEndSepPunct{\mcitedefaultmidpunct}
{\mcitedefaultendpunct}{\mcitedefaultseppunct}\relax
\EndOfBibitem
\end{thebibliography}

\pagebreak
\widetext
\begin{center}
\textbf{Supplementary Information: Superresolution Microscopy of the Volume Phase Transition of pNIPAM Microgels}
\end{center}
\setcounter{equation}{0}
\setcounter{figure}{0}
\setcounter{table}{0}
\setcounter{page}{1}
\makeatletter
\renewcommand{\theequation}{S\arabic{equation}}
\renewcommand{\thefigure}{S\arabic{figure}}
\renewcommand{\bibnumfmt}[1]{[S#1]}
\renewcommand{\citenumfont}[1]{S#1}
\section{density profiles of individual microgels}Single particle density profiles and distribution of the values obtained for the core radius and the shell thickness from fits to the density profiles of individual microgels. 
\begin{figure}[H]  \centering\includegraphics[width=16cm]{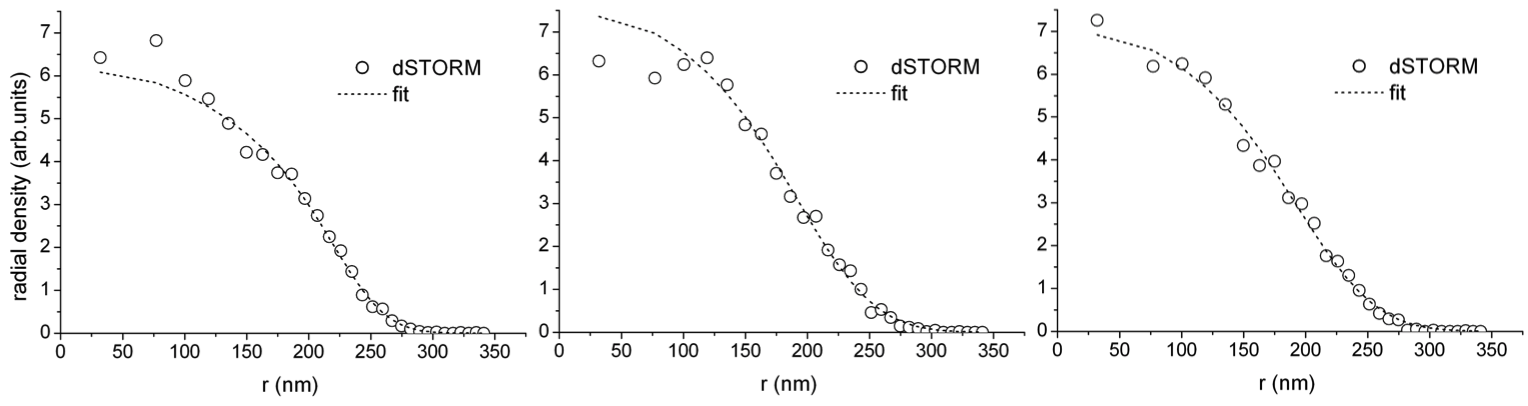}
  \caption{Measured single particle 2D density profiles $\rho_{\text{2D}}(r)$ for three different microgel particles. From left to right: R[nm]= 248, 225, 227 and 2$\sigma_{\text{surf}}$[nm]= 54, 80, 81nm.}
  \label{fgr:S1} 
\end{figure}
\begin{figure}[H] \centering\includegraphics[width=12cm]{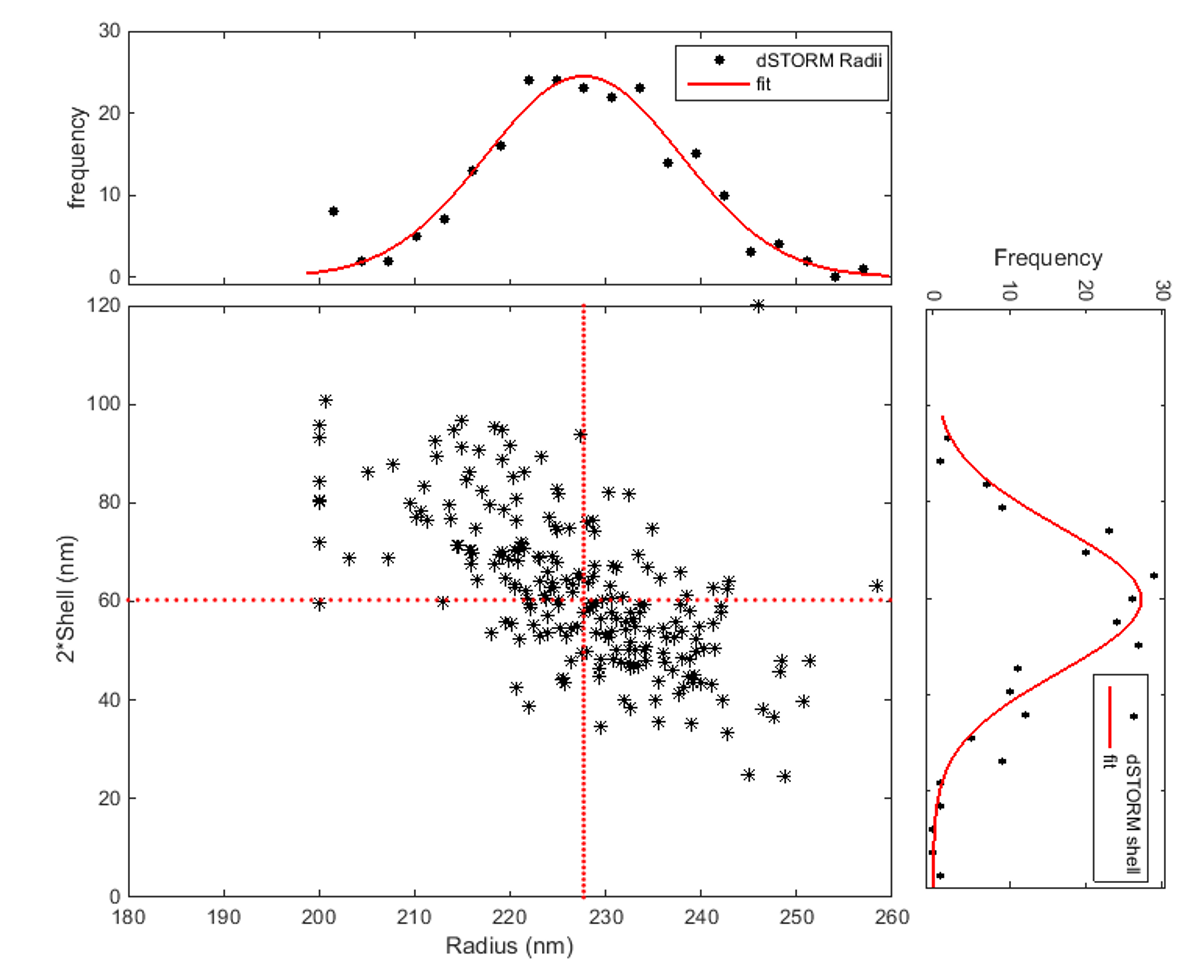}
  \caption{Characteristic density parameters derived from imaging individual microgel particles. Symbols: Values for the core radius R and the shell 2$\sigma_{\text{surf}}$ obtained from fits to the density profiles of individual microgels as shown in Fig.\ref{fgr:S1}. The distributions are shown on the top and on the right of the scatter graph. The mean core radius R = 231nm and shell thickness 2$\sigma_{\text{surf}}$  = 60nm are shown as dotted lines. The relatively large spread is due to the strong correlation between $\sigma_{\text{surf}}$ and R in the fit. The spread for the particle size R + 2$\sigma_{\text{surf}}$ is very narrow with a standard deviation of 3.1\% as shown in Fig. \ref{fgr:radprofile}(c).}
  \label{fgr:S2} 
\end{figure}

\end{document}